\documentclass{PoS}

\title{The Operator Product Expansion on the Lattice}

\ShortTitle{The Operator Product Expansion on the Lattice}

 \author{W.~Bietenholz$^a$,
 N.~Cundy$^b$, M.~G\"ockeler$^b$, R.~Horsley$^c$, H.~Perlt$^d$,
 D.~Pleiter$^a$, 
 \speaker{P.E.L.~Rakow}$^e$,%
 C.J.~Roberts$^e$, G.~Schierholz$^{a,f}$, A.~Schiller$^d$
 and J.M.~Zanotti$^c$\\
 \llap{$^a$}\ John von Neumann Institute NIC/DESY Zeuthen, 15738 Zeuthen, 
 Germany \\
 \llap{$^b$}\ Institut f\"ur Theoretische Physik, Universit\"at Regensburg, 
 93040 Regensburg, Germany\\
 \llap{$^c$}\ School of Physics, University of Edinburgh,
 Edinburgh EH9 3JZ, UK\\
 \llap{$^d$}\ Institut f\"ur Theoretische Physik, Universit\"at Leipzig, 
       PF 100 920, 04009 Leipzig, Germany\\
 \llap{$^e$}\ Theoretical Physics Division, Department of Mathematical
 Sciences, University of Liverpool, Liverpool L69 3BX\\
 \llap{$^f$}\ Deutsches Elektronen-Synchrotron DESY, 22603 Hamburg, Germany \\
        E-mail: \email{rakow@amtp.liv.ac.uk}}

\abstract{ We investigate the Operator Product Expansion (OPE) on the
 lattice by directly measuring the product $\langle J_\mu J_\nu \rangle$
 (where $J$ is the 
 vector current) and comparing it with the expectation values of 
 bilinear operators. This will determine the Wilson coefficients in the 
 OPE from lattice data, and so give an alternative to the conventional
 methods of renormalising lattice structure function calculations.
 It could also give us access to higher twist quantities such as the
 longitudinal structure function $F_L = F_2 - 2 x F_1$. We use overlap
 fermions because of their improved chiral properties, which reduces
 the number of possible operator mixing coefficients.}

\FullConference{The XXV International Symposium on Lattice Field Theory\\
		 July 30 - August 4 2007\\
		 Regensburg, Germany}

 \newcommand{\psib}{\overline{\psi}}

\begin{document}

\section{ Introduction }


   Our main theoretical tool for interpreting hadronic Deep Inelastic
 Scattering (DIS) is the Operator Product Expansion (OPE). 
 This relates the experimentally measurable electromagnetic tensor
 $W_{\mu \nu}$, 
  \begin{equation}
   W_{\mu \nu}(p,q)
   \equiv  \langle \psi(p) | J_\mu(q) J^\dag_\nu(q) | \psi(p) \rangle
  \end{equation}   
 to a sum over matrix elements of local operators. 
  \begin{equation}
   W_{\mu \nu}(p,q)
   =  \langle \psi(p) | J_\mu(q) J^\dag_\nu(q) | \psi(p) \rangle
  =  \sum_m C_{\mu \nu}^m(q) \langle \psi(p) | {\cal O}^m| \psi(p) \rangle \;. 
  \end{equation}   
  The local operators have interpretations in terms of the target hadron's
 internal structure. 

 In each term of the OPE the scales separate. 
  Dependence on the photon scale { $q$} is in the Wilson
 coefficient { $ C_{\mu \nu}^m(q) $}  while
 dependence on the quark momentum $ p $ is in the matrix
 element { $ \langle \psi(p) | {\cal O}^m | \psi(p) \rangle $}.

    There has been a long history of lattice calculations of 
 the hadronic matrix elements which occur in the OPE, but the 
 Wilson coefficients are usually calculated perturbatively. In this work
 we investigate the possibility of also calculating the Wilson
 coefficients on the lattice by looking at the product of two 
 electromagnetic currents. 

    The Wilson coefficients are independent of the target. In our
 calculation we measure the current product (Compton amplitude) 
 between quark states. We then plan to use the resulting Wilson
 coefficients together with lattice data on nucleon matrix 
 elements to look at deep inelastic scattering. 

    If we can measure the coefficients accurately enough we could 
 learn something about higher twist effects, and non-leading amplitudes
 such as the
 longitudinal structure function $F_L = F_2 - 2 x F_1$. 
 To calculate power corrections of this type we need to know both
 the matrix elements and the Wilson coefficients beyond perturbation
 theory~\cite{Martinelli}. 
 
   In this work we report on an ongoing study of the lattice OPE
 using overlap fermions. These have the
 advantage of better chiral properties, which reduces the problems
 of operator mixing, and makes $O(a)$ improvement much simpler. 
 Our earlier study using Wilson fermions 
 was described in~\cite{lat98}

\section{ Symmetry }

   If { $ p \ll q $} we will be able to truncate our set of operators
 according to their dimension. The operators we consider are the
 quark bilinears with up to three derivatives. These are the 
 operators
 $ \psib \Gamma \psi ,\;  \psib \Gamma  D_{\mu_1} \psi ,\;
 \psib \Gamma  D_{\mu_1}  D_{\mu_2}  \psi ,\;
 \psib \Gamma  D_{\mu_1}  D_{\mu_2}   D_{\mu_3} \psi $ 
 where the matrix $\Gamma$ can be any of the 16 matrices in the Clifford
 algebra. This means that there are a total of 
 $16 \times (1 + 4 + 4^2 + 4^3 ) = 1360 $ operators to consider. 
 Do we need to find 1360 different $C^m$ values? 

    To reduce the number of independent coefficients
 we want to choose a $q$ vector with as much lattice symmetry as possible,
 so we have taken $ q \propto (1, 1, 1, 1) $. The data presented here
  are for the choice  $ aq = \left( \frac{\pi}{2} ,  \frac{\pi}{2},
  \frac{\pi}{2} ,  \frac{\pi}{2} \right) $. 

  In the expansion for $W_{44}$ we know that rotations and reflections
 that mix the space direction $x, y, z$ are symmetries. So it
 is obvious that the Wilson coefficients of
 $ \psib \gamma_1  D_1 \psi , \; 
 \psib \gamma_2  D_2 \psi$  and
 $\psib \gamma_3  D_3 \psi $  
 should all be the same. 
   Exploiting symmetry this way reduces the original $1360$ coefficients
 down to only $ 67$, a much more manageable problem. 
  We can also use symmetries to relate different components of
 $W_{\mu \nu}$. For example, the Wilson coefficient of 
 $ \psib \gamma_4  D_4 \psi$  in $W_{44} $
 is the same as the $C$ for
  $ \psib \gamma_3  D_3 \psi $ in $W_{33} $. 

\section{ Lattice Details }
  
  We are carrying out calculations with overlap valence fermions,
 due to their superior chiral properties. However because of the 
 high cost of a full dynamical overlap calculation, we have to use
 gauge configurations calculated with $N_f=2$
 dynamical clover fermions. 

   The overlap fermions are calculated with $\rho = 1.5 $
 and bare mass $ a m = 0.024$. 
 The results discussed in this work  used a 
 $16^3 \times 32$ lattice, simulated with $N_f=2$ clover fermions, 
 at $\beta = 5.29$ and $\kappa = 0.1350$, 
 which corresponds to a lattice spacing of $a = 0.075 fm$.  

  The Green's functions have been $O(a)$ improved, which is easy to
 do with overlap fermions. The current $J_\mu(x)$ was represented by
 the local current  $\psib(x) \gamma_\mu \psi(x)$. Since we are measuring
 operators between quark states we have to fix the gauge. We used 
 the lattice Landau gauge. We use a momentum source~\cite{momsource,ZNP}
 for all Greens functions, which leads to a great reduction in statistical 
 noise. 

     We used a momentum transfer
  $ aq = \left( \frac{\pi}{2} ,  \frac{\pi}{2},
 \frac{\pi}{2} ,  \frac{\pi}{2} \right) $  corresponding to 
 $|q| = 8.3 GeV$. For this $q$ value we measured the two-point and
 three-point functions for 28 different $p$ vectors, with a large spread in
 directions. Here we give results for $\langle J_4 J_4 \rangle$, we are
 also collecting data on other components of the current-current tensor. 

   We only consider the flavour non-singlet case, so we do
 not include any purely gluonic operators in our calculation. 

 \section{Strategy}

  We calculate the Compton scattering amplitude for a quark with 
 a large number of $p$ values. The result is a Dirac matrix for each $p$
 vector, i.e. 16 complex numbers. So our data on the
 Compton amplitude consist of $16 M_p $ complex numbers. 
   We also calculate the operator Greens functions for each of
 our operators, for all of these $p$ values. This gives
  $(16 M_p) \times N_O$  numbers as our data on the operators. 

    The information we want to extract from all this data are 
  the  $N_O$ Wilson coefficients which best reproduce
 the Compton scattering amplitudes. 

   This is essentially a linear algebra problem, and is 
 best written as a matrix equation  
 \begin{equation} 
  W_{\mu \nu}(p_i,q) = \sum_m 
 \langle \psi(p_i) | {\cal O}^m| \psi(p_i) \rangle\;  C_{\mu \nu}^m(q) 
 \end{equation}

\begin{equation}
\left(
\begin{array}{ccc}
W^{p_1}  \\
\vdots    \\[-0.2em]
\vdots    \\
W^{p_M}  \\
\end{array}
\right) 
=
\left(
\begin{array}{ccc}
 O_{1}^{p_1}  &  \hspace*{-0.cm}\cdots\hspace{-0.cm} &
 O_{N}^{p_1}   \\
\vdots   &  & \vdots \\[-0.2em]
\vdots   &  & \vdots \\
 O_{1}^{p_M}  & \hspace*{-0.cm}\cdots\hspace*{-0.cm} &
 O_{N}^{p_M}   \\
\end{array}
\right)
\left(
\begin{array}{ccc}
C_{1}  \\
\vdots \\
C_{N}  \\
\end{array}
\right)\;.
\label{matrix}
\end{equation}

 $$ (16 M_p) \quad \qquad (16 M_p \times N_O) \cdot (N_O)  $$

 \section{Singular Value Decomposition (SVD)}

    There are two difficulties with this system of equations. Firstly
 if $16 M_p > N_O$ the system is overdetermined, so no exact solution
 will be possible. Secondly, some of the operators might be linearly
 dependent, in which case the system is also ill-conditioned,
 and there will not even be a unique best approximation to the solution.

 Nevertheless by using Singular Value Decomposition (SVD)~\cite{recipe},
 the standard technique for problems of this type, we can find
 values of the coefficients $C_i$ which give a very good, and very stable,
 approximate solution.

    We can always factorise the operator matrix $O$ as
 { \begin{eqnarray}
      O &=& U \omega V^T \\
  O_{i,j} &=& \sum_{k=1}^N U_{ik} \omega_k V_{jk}
 \end{eqnarray} }
 where $U$ and $V$ are column-orthonormal, $U$ is $(16 M_p) \times N_O$,
 $V$ is $N_O \times N_O$ and $\omega$ is a diagonal $N_O \times N_O$ matrix,
 with positive real eigenvalues $\omega_k$ arranged in descending order.
 The $\omega_k$ are the analogues of eigenvalues for a rectangular
 matrix, while the matrices $U$ and $V$ contain the eigenvectors for the
 system. 
 If some of the $\omega_k$ are very small they can safely be dropped
 from the sum without significantly changing $O$. 

 We define $O^{(n)}$
 as the approximation to $O$ that we get by keeping the $n$ largest
 $\omega$ values, and dropping the $N-n$ smallest values.
 This gives
 \begin{equation}
  O^{(n)}_{i,j} = \sum_{k=1}^n U_{ik} \omega_k V_{jk}
      \equiv U^{(n)} \omega^{(n)} (V^{(n)})^T
 \end{equation}
 where now $U^{(n)}$ is $(16 M_p) \times n$, $\omega^{(n)}$ is $n \times n$
 and  $V^{(n)}$ is $N_O \times n$. $O^{(n)}$ is still $(16 M_p) \times N_O$.
 It can be shown that the least-squares solution to
 $ O^{(n)} C = W $ is
 $ C = V^{(n)} (\omega^{(n)})^{-1} ( U^{(n)})^T W $. 

   What singular value decomposition has done is to find the $N-n$
 linear combinations of our original operators which have the least
 influence on $W$, and discarded those combinations. Discarding 
 these operators makes the problem much more stable. However if we
 discard too many operators, $O^{(n)}$ will no longer be a good
 approximation  to $O$, and so the solution of the approximate
 system will not give a good approximate solution to the true 
 problem. 

\begin{figure}[htb] \begin{center}
 \includegraphics[angle=270,width=0.7\textwidth]{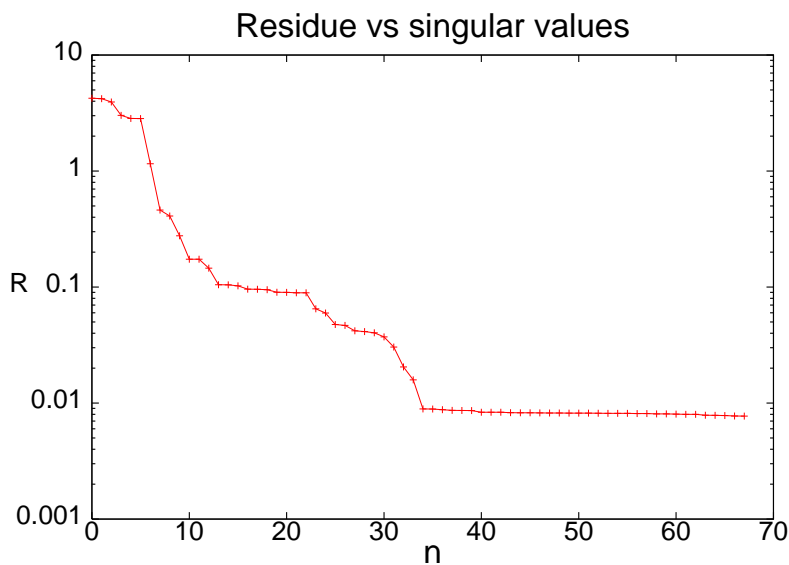}
 \end{center} 
 \caption{The residue of the fit as a function of $n$, the number
 of singular values retained. \label{Figresid}} 
   \end{figure}

    To solve a system by SVD we vary $n$, the number of $\omega_k$ values
 retained, looking for a region where the residue 
 $R = (W - O C)^2$ is small and the system is stable. 
 Fig.\ref{Figresid} shows how the residue declines as the number
 of singular values is increased. 
  A plateau is reached beyond $n \approx 40$. Beyond this point
 adding more operators does not decrease the residue significantly. 

    To judge the stability of the fit, we look at the value of one
 of the Wilson coefficients, and see how it depends on $n$. 
 In Fig.{\ref{FigCval} we show the result for the choice $C_2$ in our list
 of operators, the operator 
 $\psib ( \gamma_1  D_1 + \gamma_2  D_2 + \gamma_3  D_3 )\psi$. 
 At first the coefficient changes dramatically as operators are added, 
 but by the time $n$ has reached 40, there are only minor changes. 
 If $n$ is made too large, there is a risk that we will start ``fitting
 to the noise'', and the value of the coefficient will become noisy. 
 There are indeed some fluctuations beyond $n=50$, but they are not
 unduly large. 

 \begin{figure}[htb]
 \begin{center}
 \includegraphics[angle=270,width=0.7\textwidth]{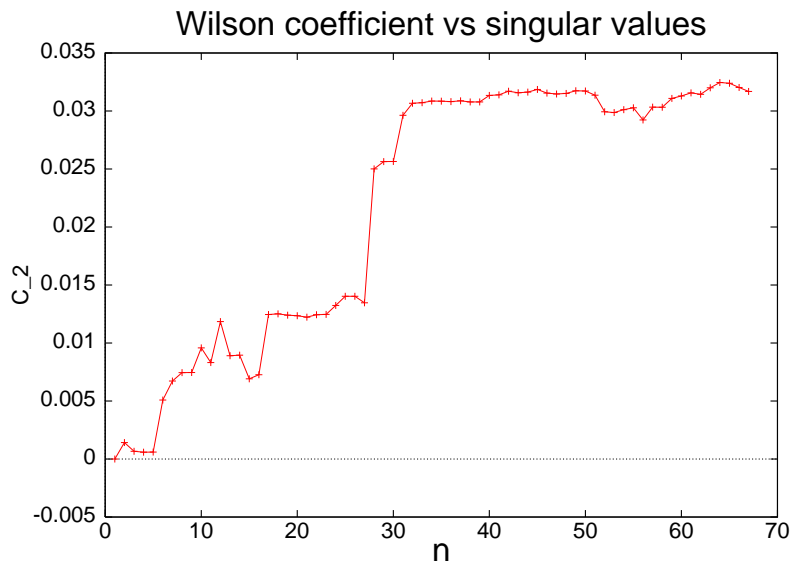}
 \end{center}
 \caption {The value of $C_2$, the Wilson coefficient for 
 ${\psib} ( \gamma_1 {D}_1
 + \gamma_2 {D}_2 + \gamma_3 {D}_3 )\psi$, 
 as a function of $n$, the number
 of singular values retained. \label{FigCval} }
 \end{figure}

   Another way of judging the quality of the fit is to exclude one
 momentum value from the fit, and see how well it is predicted by 
 the data at all the other momentum values. The result is shown 
 in Fig.\ref{predict}. Again we see that we need at least 40
 singular values to produce a fit with good predictive power. 
 If, at large $n$ we started fitting to the errors in the data, 
 we would expect to see the predictive power becoming worse. 
 There doesn't seem to be any sign of this happening. 

 \begin{figure}[htb]
 \begin{center}
 \includegraphics[angle=270,width=0.7\textwidth]{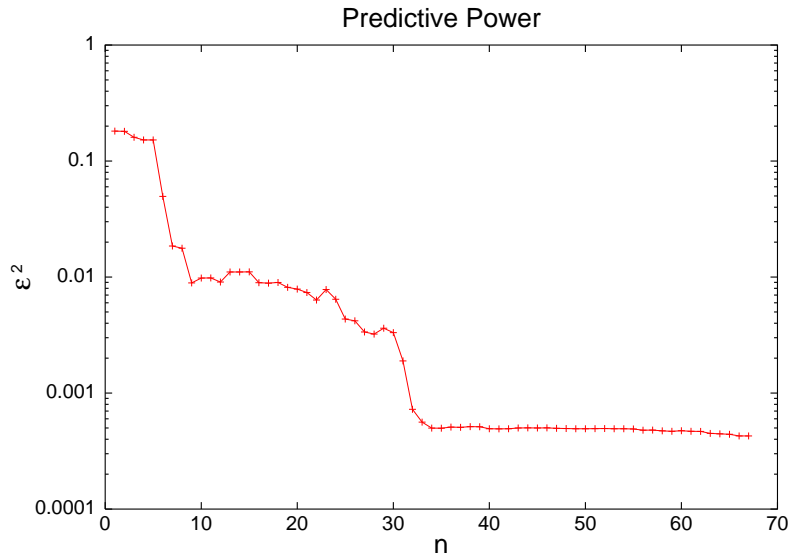}
 \end{center}
  \caption{ The accuracy with which the fit predicts the value
 of $W_{\mu \nu}$ as a function of $n$. \label{predict} }
 \end{figure} 

 \section{Results}


   In Fig.\ref{chiralfig} we show the results of our fit. 
   Chiral symmetry shows up well. Operator number 1, the
 operator $\psib {\bf 1}\psi$, 
 and operators 7 to 16, which are two-derivative operators
 proportional to the unit matrix or to $\sigma$ matrices, 
 are ruled out by chiral symmetry, so they would have 
 Wilson coefficients of 0 in the chiral limit. Here, 
 using overlap fermions, we find that these coefficients are 
 indeed small. This contrasts with earlier work with Wilson
 fermions,~\cite{lat98},~\cite{lat06}, where the operator
 $\psib {\bf 1} \psi$ was prominent. 

 \begin{figure}[htb]
 \begin{center}
 \includegraphics[angle=270,width=0.7\textwidth]{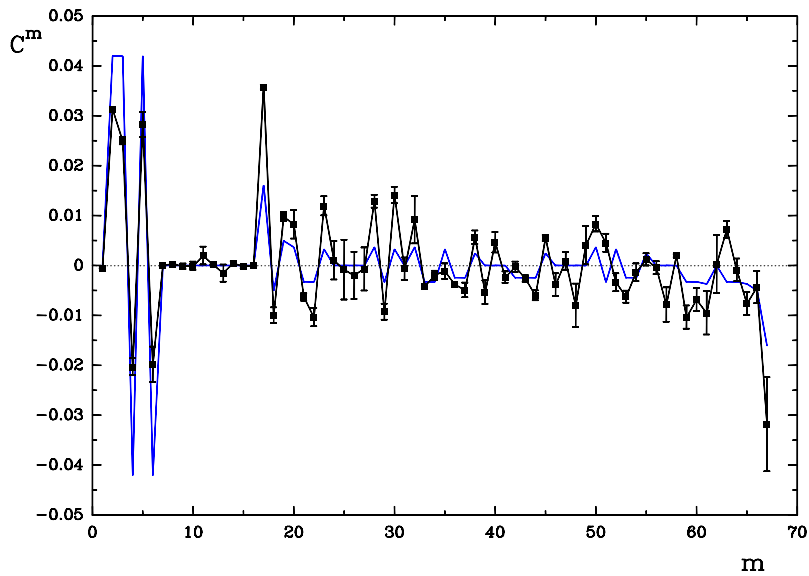}
 \end{center}
 \caption{ The Wilson coefficients for $W^{44}$ determined on the 
 lattice according to our procedure.  They are compared with the 
 lattice tree-level values, shown by the blue line. 
 \label{chiralfig} }
 \end{figure}

    We have calculated the tree-level Wilson coefficients for overlap
 fermions, following the calculation set out in~\cite{lat06}. The
 tree-level results are shown by the blue line in Fig.\ref{chiralfig}. 
 We see that the pattern of Wilson coefficients is very similar. 

     These results are at a single value of $|a q|$. We intend to
 gather data at other values to investigate the size of lattice artefacts,
 which should be $O(a^2)$ for overlap fermions.

\section{Conclusions}

   Using Singular Value Decomposition we have been able to reconstruct
 Wilson coefficients from lattice data on the electromagnetic tensor. 
 Statistical errors are small, due to our use of momentum sources
 for the inversions. 

     The results we have look reasonable, they follow a pattern similar
 to that seen at tree-level, and they show the effects expected 
 from chiral symmetry. So far we only have data from a rather large
 value of $|a q|$, we plan to look at more $q$ values to check for
 $O(a^2)$  lattice errors. 

\section{Acknowledgements}
   
    The calculations for this work were carried out on an 
 IBM pSeries 690 in Berlin, belonging to the HLRN.



\begin{thebibliography}{99}

 \bibitem{Martinelli}
  G.~Martinelli and C.~T.~Sachrajda,
  Nucl.\ Phys.\  B {\bf 478} (1996) 660
  [arXiv:hep-ph/9605336].

\bibitem{lat98}
  S.~Capitani, M.~G\"ockeler, R.~Horsley, H.~Oelrich,
 D.~Petters, P.E.L.~Rakow and G.~Schierholz,
  Nucl.\ Phys.\ Proc.\ Suppl.\  {\bf 73} (1999) 288
  [arXiv:hep-lat/9809171].
 
\bibitem{momsource}
  M.~G\"ockeler, R.~Horsley, H.~Oelrich, H.~Perlt,
 P.~Rakow, G.~Schierholz and A.~Schiller,
  Nucl.\ Phys.\ Proc.\ Suppl.\  {\bf 63} (1998) 868
  [arXiv:hep-lat/9710052].

\bibitem{ZNP}
 M. G\"ockeler, R. Horsley, H. Oelrich, H. Perlt, D. Petters,
 P.E.L.~Rakow, A. Sch\"afer, G. Schierholz and A. Schiller,   
  Nucl.\ Phys.\  B {\bf 544} (1999) 699
  [arXiv:hep-lat/9807044].

 \bibitem{recipe} W.H.~Press  S.A.~Teukolsky, W.T.~Vetterling
 and B.P.~Flannery, {\it Numerical Recipes},
 Cambridge University Press, Cambridge (1989).

\bibitem{lat06}
  M.~G\"ockeler, R.~Horsley, H.~Perlt, P.E.L.~Rakow,
 G.~Schierholz and A.~Schiller [QCDSF Collaboration],
  PoS {\bf LAT2006} (2006) 119 [arXiv:hep-lat/0610064].

\end{thebibliography}
\end{document}